\documentclass[aps,showpacs,preprintnumbers,amsmath, amssymb]{revtex4}

\usepackage{float}
\usepackage{graphics,epsfig}
\usepackage{graphicx}
\usepackage{dcolumn}
\usepackage{bm}

\begin{document}
\baselineskip=0.8 cm

\title{{\bf Perturbations around the AdS Born-Infeld black holes}}

\author{Yunqi Liu, Bin Wang}
\affiliation{ INPAC and Department of Physics,
Shanghai Jiao Tong University, Shanghai 200240,
China}
\vspace*{0.2cm}
\begin{abstract}
\baselineskip=0.6 cm
\begin{center}
{\bf Abstract}
\end{center}
We study the quasinormal mode of the perturbation
of the scalar field interacting with the
electromagnetic field in the backgrounds of
Schwarzschild AdS black hole, the
Reissner-Nordstrom AdS black hole and the
Born-Infeld AdS black hole. We disclose influence
by the electric charge $Q$, the coupling between
the scalar and electromagnetic field $q$ and the
Born-Infeld parameter $b$ on the quasinormal
frequencies of the scalar perturbation. We
observe the possible growing mode in the
perturbation when the scalar field strongly
couples to the electromagnetic field. When the
electromagnetic field becomes nonlinear described
by the Born-Infeld electrodynamics, the
nonlinearity described by the Born-Infeld
parameter can hinder the growing mode to appear
in the scalar perturbation. The quasinormal mode
of the scalar field perturbation can help us
further understand the physics in the holographic
superconductor.
\end{abstract}

\maketitle
\newpage
\vspace*{0.2cm}

\section{Introduction}

Quasinormal mode (QNM) describes the
perturbations in the surrounding geometry of a
black hole. Its frequency is entirely fixed by
the black hole parameters. The QNM is believed as
a unique fingerprint of the black hole existence
(for reviews on this topic, see
\cite{Nollert,Kokkotas,profWang,roman} and
references therein). In the last decade, QNM has
become an intriguing subject. It can serve as a
testing ground of fundamental physics. It is
widely believed that the study of QNM can help us
get deeper understandings of the
AdS/CFT\cite{profWang, roman,
Horowitz,wang,wang1}, dS/CFT \cite{Abdalla}
correspondences, loop quantum gravity \cite{Hod},
the phase transition of black holes
\cite{Koutsoumbas} and the extra dimension
\cite{Ruffini} etc.

In most studies of the QNM, the wave dynamics of a single classical
field propagating outside the black hole is taken into account. For
a stable black hole background, the perturbation of a single
classical field will die out finally due to the black hole
absorption. Recently more interest has been focused on the strongly
coupled theory by considering the Einstein-Maxwell field interacting
with a charged scalar field. It has been shown that the bulk AdS
black hole can become unstable and scalar hair can condense below a
critical temperature. In the light of AdS/CFT correspondence, it was
realized that the instability of the bulk black hole corresponds to
a second order phase transition from normal state to superconducting
state which brings the spontaneous $U(1)$ symmetry breaking
\cite{Hartnoll}. Along this line, there have been a lot of
investigations concerning the application of AdS/CFT correspondence
to condensed matter physics by considering interactions among
classical fields \cite{Herzog}-\cite{panwang1}. It was found that
the QNM can be an effective tool to disclose the second order
transition between a non-superconducting state at high temperatures
and a superconducting state at low temperatures.

In this work we will consider the nonlinear
extension on the gauge fields in Einstein-Maxwell
theory and investigate the perturbation of the
scalar field in interaction with the nonlinear
electromagnetic field. Recently in the bulk
Schwarzschild AdS  black hole and the AdS
Gauss-Bonnet black hole backgrounds, the effect
of the Born-Infeld electrodynamics on the
holographic superconductor has been investigated
in \cite{chen}. To see exactly how the higher
derivative corrections to the gauge matter fields
influence the perturbation around the bulk black
hole and make the scalar field condense, we are
going to examine the QNM of the scalar
perturbation when it is coupled with the
Born-Infeld electromagnetic field. In our
investigation we will adopt the AdS Schwarzschild
and the AdS Born-Infeld black hole solutions as
backgrounds.

The structure of the paper is as follows: In
Sec.II, we will introduce the action and deduce
the Einstein equation and equations of motions
describing the nonlinear electromagnetic field
and the charged scalar perturbation in the
five-dimensional AdS Schwarzschild and
Born-Infeld black hole spacetimes. In Sec.III, we
will present the numerical results and disclose
the properties of the QNM of the scalar
perturbation when it is coupled with the
nonlinear electromagnetic field. We will
summarize our results in Sec.IV.

\section{Action and Basic Equations}

The general action describing the
Einstein-Born-Infeld field interacting with a
charged scalar field in the five-dimensional AdS
background has the form
\begin{eqnarray}\label{action}
S=\frac{1}{2\kappa^2}\int d^5 x\sqrt{-g}(R+\frac{12}{l^2})+\int d^5
x\sqrt{-g}\left[\frac{1}{b}\left(1-\sqrt{1+\frac{b
F^{ab}F_{ab}}{2}}\right)-|\nabla\Psi-i q A\Psi
|^2-m^2|\Psi|^2\right],
\end{eqnarray}
where $\kappa$ is the five-dimensional
gravitational constant $\kappa^2=8 \pi G_5$, and
$G_5$ is the five-dimensional Newton
constant,~$g$ is the determinant of the
background black hole metric, ~$R$ is the Ricci
curvature,~$l$ describes the AdS radius, $q$ is
the coupling between the electromagnetic field
and the scalar field, and $m$ indicates the mass
of the scalar field. Here $b$ is the Born-Infeld
parameter. In the limit $b\rightarrow 0$, the
Born-Infeld electrodynamics reduces to the
Maxwell theory.

We assume a static black hole with the metric
\begin{eqnarray}\label{ansatz}
ds^2=-f(r)e^{-\chi(r)}dt^2+\frac{dr^2}{f(r)}+\frac{r^2}{l^2}d\Omega_3
^2~.
\end{eqnarray}
where
\begin{eqnarray}
d\Omega_3 ^2=d\theta^2+sin^2 \theta(d\phi^2+sin^2
\phi d\psi^2).\nonumber
\end{eqnarray}
Its Hawking temperature, which will be
interpreted as the temperature of the CFT, has
the form
\begin{eqnarray}\label{temperature}
T=\left.\frac{f'(r)e^{-\chi(r)/2}}{4 \pi}\right|_{r=r_h}~.
\end{eqnarray}
where $r_h$ is the radius of the black hole
horizon.

For a static solution, we assume that the
electromagnetic field and the scalar field take
the forms
\begin{eqnarray}
A_t=\phi(r)dt,~~~~\Psi=\Psi(r).\nonumber
\end{eqnarray}

Varying the action (\ref{action}) with respect to
the metric $g_{\mu \nu}$, we can obtain the
Einstein equations
\begin{eqnarray}\label{Einsteineqn0}
0&=&\chi '(r)+\frac{4}{3} \kappa^2 r \left(\frac{e^{\chi(r)} q^2 \phi(r)^2 \Psi(r)^2}{f(r)^2 }+\Psi'(r)^2\right),\nonumber\\
0&=&f'(r)-\frac{2}{r}+\frac{2f(r)}{r}-\frac{4r}{l^2}+\frac{2}{3}
\kappa ^2 r \left[m^2 \Psi(r)^2+\frac{e^{\chi (r)} q^2 \phi(r)^2
\Psi(r)^2}{f(r)}+\frac{1}{b}\left[\left(1-b
\phi'(r)^2\right)^{-1/2}-1\right]\right].
\end{eqnarray}
The prime denotes the derivative with respect to
$r$.

In the probe limit when $\kappa^2\rightarrow 0$, the solution of the
Einstein equation gives the Schwarzschild AdS black hole with
\begin{eqnarray}\label{sads}
f(r)=1+\frac{r^2}{l^2}-\frac{r_h^2}{r^2}\left(1+\frac{r_h^2}{l^2}\right),\nonumber
\end{eqnarray}
where  $r_h$ is the black hole horizon obtained from $f(r_h)=0$.

For the nonzero parameter $\kappa^2$, we take the
backreaction of classical fields onto the
spacetime into account. In studying the
perturbation\textbf of the massless scalar field
$\Psi(r)$ around the black hole, we can reduce
Eq.(\ref{Einsteineqn0}) into
\begin{eqnarray}\label{Einsteineqn}
0&=&\chi '(r),\nonumber\\
0&=&f'(r)-\frac{2}{r}+\frac{2f(r)}{r}-\frac{4r}{l^2}+\frac{2}{3b}
\kappa ^2 r \left[\left(1-b
\phi'(r)^2\right)^{-1/2}-1\right].
\end{eqnarray}
If we know the behavior of the electromagnetic
field, we can obtain the black hole metric from
the above Einstein equations.

The equation governing electromagnetic field can
be obtained by varying the action (\ref{action})
with respect to $\phi$, which reads
\begin{eqnarray}\label{Einsteineqn1}
0&=&\left[\phi ''(r)+\left(\frac{\chi'(r)}{2
}+\frac{3}{r}\right)\phi '(r)\right](1-b e^{\chi (r)}\phi '(r)^2
)\nonumber\\&& +\frac{b}{2} \phi '(r) e^{\chi (r)}\left(\chi'(r)
\phi'(r)^2+2 \phi'(r) \phi''(r)\right)-\frac{2 q^2 \phi(r) \Psi(r)^2
}{f(r) } \left(1-b e^{\chi(r)} \phi'(r)^2\right)^{3/2}.
\end{eqnarray}
Considering the perturbation of the massless scalar field $\Psi(r)$,
Eq.(\ref{Einsteineqn1}) reduces to
\begin{eqnarray}\label{phieqn}
0&=&\left[\phi ''(r)+\left(\frac{\chi'(r)}{2
}+\frac{3}{r}\right)\phi '(r)\right](1-b e^{\chi (r)}\phi '(r)^2 )
+\frac{b}{2} \phi '(r)^2 e^{\chi (r)}\left(\chi'(r)
\phi'(r)+2\phi''(r)\right).
\end{eqnarray}

Doing the rescales
\begin{eqnarray}\label{symmetry}
&e^{\chi} \rightarrow a^{2} e^{\chi},~\phi \rightarrow a^{-1}\phi,~t
\rightarrow a t,~\omega\rightarrow a^{-1} \omega.\nonumber\\
&r\rightarrow r/l,~q \rightarrow q
l,~\omega\rightarrow l
\omega,~\phi\rightarrow \phi/ l,\nonumber\\
&q \rightarrow aq,~\phi \rightarrow
\phi/a,~\kappa^2 \rightarrow \kappa^2
a^{2},~b\rightarrow a^2 b,
\end{eqnarray}
where  the first symmetry guarantees $\chi(r)=0$,
the second symmetry sets AdS radius as unity, the
third one relates the backreaction to the charge
of the scalar field and we can set $\kappa=1$
without loss of generality when the backreaction
is not null.

Substituting $\chi(r)=0 $ , Eq.(\ref{phieqn})
becomes
\begin{equation}\label{phieq2}
0=b \phi'(r)^2 \phi''(r)+\left(1-b
\phi'(r)^2\right) \left(\frac{3
\phi'(r)}{r}+\phi''(r)\right).
\end{equation}
We can solve this equation of motion of the
electromagnetic field analytically, which reads
\cite{cai}
\begin{equation}\label{phiso2}
\phi(r)=U+\frac{ Q
}{r^{2}}_{2}F_{1}\left[\frac{1}{3},\frac{1}{2},\frac{4}{3},-\frac{4bQ^2}{r^6}\right],
\end{equation}
where $Q$ and $U$ are integral constants, $U$ can
be eliminated by the constraint at the horizon,
$\phi(r_h)=0$,
\begin{equation}
U=-\frac{1}{r_h^2}
_{2}F_{1}\left[\frac{1}{3},\frac{1}{2},\frac{4}{3},-\frac{4bQ^2}{r^6}\right].\nonumber
\end{equation}
At infinity, Eq.(\ref{phiso2}) asymptotically
behaves as
\begin{equation}
\phi(r)\sim U-\frac{Q}{ r^2}.\nonumber
\end{equation}

In the limit $b\rightarrow0$, Eq.(\ref{phieq2})
is reduced to
\begin{equation}\label{phieq3}
0=\frac{3 \phi'(r)}{r}+\phi''(r)
\end{equation}
which is the equation of motion of the
electromagnetic field in the Maxwell theory. The
solution  reads
\begin{eqnarray}
\phi(r)=\frac{Q}{r_h^2}-\frac{Q}{ r^2}.\nonumber
\end{eqnarray}

Substituting  Eq.(\ref{phiso2}) into
Eq.(\ref{Einsteineqn}), the differential equation
of $f(r)$ is rewritten as
\begin{equation}\label{ffun}
0=f'(r)+\frac{2 f(r)}{r}-\frac{2}{r}-\frac{4
r}{l^2}-\frac{2 \kappa^2 r}{3 b}+\frac{2
\kappa^2}{3br^2}\sqrt{4 b Q^2+r^6},
\end{equation}
which has an analytical solution
\begin{equation}\label{biads}
f(r)=1+\frac{P}{r^2}+\frac{r^2}{l^2}+\frac{\kappa^2
r^2}{6b}-\frac{2\kappa^2}{3br^2}\int\sqrt{4bQ^2+r^6}dr.
\end{equation}
After the integration, we obtain
\begin{equation}\label{biads}
f(r)=1+\frac{P}{r^2}+\frac{r^2}{l^2}+\frac{\kappa^2
r^2}{6b}-\frac{\kappa^2}{6br}\sqrt{4bQ^2+r^6}+\frac{\kappa^2Q^2}{r^4}_{2}F_{1}\left[\frac{1}{3},\frac{1}{2},\frac{4}{3},-\frac{4bQ^2}{r^6}\right],
\end{equation}
where $P$ is the integration constant determined
by the boundary condition at the horizon $r_h$,
\begin{equation}
P=-r_h^2\left(1+\frac{
r_h^2}{l^2}\right)-\frac{\kappa^2
r_h^4}{6b}+\frac{\kappa^2
r_h^4}{6b}\sqrt{r_h^6+4bQ^2}-\frac{\kappa^2Q^2}{r_h^2}
_{2}F_{1}\left[\frac{1}{3},\frac{1}{2},\frac{4}{3},-\frac{4bQ^2}{r^6}\right].
\end{equation}
Eq.(\ref{biads}) is exactly the solution of the
five-dimensional Born-Infeld-Anti-de
sitter(BIAdS) black hole described in \cite{cai}.
Taking the limit $b\rightarrow0$, Eq.(\ref{ffun})
changes into
\begin{eqnarray}
0&=&f'(r)+\frac{2 f(r)}{r}-\frac{2}{r}-\frac{4
r}{l^2}+\frac{4 \kappa^2 Q^2}{3 r^5}
\end{eqnarray}
which gives the Reissner-Nordstrom AdS black hole
with the metric coefficient
\begin{eqnarray}\label{rnads}
f(r)&=&1+\frac{r^2}{l^2}+\frac{2 \kappa^2
Q^2}{3r^4}\left(1-\frac{r^2}{r_h^2}\right)-\frac{r_h^2}{r^2}\left(1+\frac{r_h^2}{l^2}\right).
\end{eqnarray}
Thus in contrast to the bakcground in the probe limit
Eq.(\ref{sads}), we can see that the backreaction of the
electromagnetic field on the Einstein gravity makes the black hole
charged.

Now we consider the scalar field perturbing around the AdS black
hole background (\ref{ansatz}), we have the wave equation of the
charged scalar field directly from the action
\begin{eqnarray}\label{psieqn}
0&=&\psi ''(r)+\psi '(r)
\left[\frac{3}{r}+\frac{f'(r)}{f(r)}-\frac{\chi
'(r)}{2}\right]+\psi(r) \frac{(\omega +q
\phi(r))^2 e^{\chi (r)}}{f(r)^2}~,
\end{eqnarray}
where we have separated the radial part of the scalar field into
$\Psi(r)=\psi(r) e^{-i \omega t} $. Here $\omega$ indicates the
frequency of the perturbation.

Near the black hole horizon $r\sim r_h$, we can
impose the incoming wave boundary condition
\begin{equation}\label{incoming}
\psi(r)\sim(r-r_h)^{- i\frac{\omega}{4\pi T}}.
\end{equation}
Introducing a new variable $\varphi$, we separate
$\psi(r)=\mathcal{\Re}(r)\varphi(r)$ where
$\mathcal{\Re}(r)=exp[-i\int^{r}_{r_h}\frac{\omega+q\phi(r)}{f(r)}]$,~which
asymptotically approaches  Eq.(\ref{incoming}) at
the horizon. Then at the horizon
$\left.\varphi\right|_{r=r_h}=const.$ and
Eq.(\ref{psieqn}) can be expressed as
\begin{eqnarray} \label{mainequation}
0&=&\varphi ''(r)+B_1(r) \varphi '(r)+B_2(r) \varphi(r)~,
\end{eqnarray}
with
\begin{eqnarray}
B_1(r)&=&\frac{f'(r)}{f(r)}-\frac{2 i[q\phi
(r)+\omega]}{
f(r)}+\frac{3}{r}~,\nonumber\\
B_2(r)&=&-\frac{i [3(q\phi (r)+\omega)+r
\phi'(r)]}{r f(r)}~.
\end{eqnarray}

At the spatial infinity $r\sim \infty$,
$\varphi(r)$ behaves as
\begin{equation}\label{asymptotic behavior}
\varphi(r)\sim
\frac{\varphi_{-}}{r^{\lambda_{-}}}+\frac{\varphi_{+}}{r^{\lambda_{+}}}.
\end{equation}
where $\lambda_\pm=d/2\pm (d/2 +m^2 l^2
)^{1/2}=2\pm 2$ for taking $m=0, d=4$. We choose
$\varphi_-=0$ in our following discussion, so
that we can only relate the scalar operator in
the field theory dual to the branch $\varphi_+$
to describe the condensation.

We can set the boundary conditions at the horizon
\begin{eqnarray}\label{boundary}
\varphi|_{r=r_h}&=&1~,\nonumber\\
\left.\frac{\varphi^{\prime}}{\varphi}\right|_{~r=r_h}
&=&-\left.\frac{B_2(r)}{B_1(r)}\right|_{r=r_h}~
\end{eqnarray}
to solve Eq.(\ref{mainequation}).

Substituting $f(r)$ and $\phi(r)$ obtained from
solving the Einstein equation and the
electromagnetic equation into
Eq.(\ref{mainequation}), we can get the
differential equation governing the perturbation
of scalar field outside the AdS black hole. Using
the boundary conditions (\ref{boundary}) and
imposing $\varphi_-=0$ at infinity, we can
calculate the frequency $\omega$ of the  scalar
perturbation. We will carry out the numerical
computation by using the shooting method and
disclose the properties of the QNM of the
massless charged scalar perturbation when it is
coupled to the nonlinear electromagnetic field.

\section{numerical results}

In this section we present our numerical results of the scalar
perturbation. We will first focus on the perturbation in the
background of the Schwarzschild AdS black hole. When there is no
coupling between the scalar field and the electromagnetic field,
namely $q=0$, we can reproduce the QNM of the single scalar field
perturbation obtained in \cite{Konoplya}. This shows that our
shooting method is effective in the numerical calculation.

Now we turn on the coupling parameter $q$, the
quasinormal frequencies of the scalar
perturbation when it is coupled with the electric
field around the Schwarzschild AdS black hole are
listed in table I.  In obtaining the numerical
results, we have fixed the black hole size
$r_h=0.1$ and the charge of the electromagnetic
field $Q=0.1$.  When $b=0$, our result goes back
to \cite{hexi}. For other fixed nonzero
Born-Infeld parameter $b$, we see the same
qualitative feature that with the increase of the
coupling parameter $q$, both the real part and
the absolute value of the imaginary part of the
quasinormal frequencies become smaller. This
shows that when the coupling between the scalar
field and electromagnetic field is stronger, the
scalar perturbation on background spacetime
decays slower. When $q$ is large enough, the
imaginary part of the QNM can become positive
which indicates that the background spacetime
becomes unstable. To see the nonlinear effect of
the electromagnetic field, we can fix the
coupling parameter $q$ and find that with the
increase of the parameter $b$, both the real part
and the absolute value of the imaginary part of
the quasinormal frequencies become bigger. This
shows that when there is coupling between the
scalar field and the electromagnetic field, the
nonlinearity of the electromagnetic field can
influence the scalar field perturbation and make
it decay faster. Furthermore we observe that when
the Born-Infeld parameter is bigger, it needs
stronger coupling $q$ to destroy the bulk
spacetime to make it unstable.

\begin{center}\label{table4}
\begin{table}[ht]
\caption{\label{table4} The dependence of the QNM on $q$ and $b$ for
 Schwarzschild AdS BH with $r_h=0.1$, $Q=0.1$.  }
\begin{tabular}{p{10mm}|p{15mm}|p{45mm}}
\hline \hline
q& b&$\omega$~\\
[0.5ex]\hline
0.1 &0&$2.968076-0.011512i$~\\[0.5ex]
&0.001&$3.463266-0.013715i$~\\[0.5ex]
&0.002&$3.539744-0.014161i$~\\[0.5ex]
&0.003&$3.580766-0.014419i$~\\[0.5ex]
&0.004&$3.608080-0.014598i$~\\
[0.5ex]\hline
0.3 &0&$1.029025-0.002924i$~\\[0.5ex]
&0.001&$2.514249-0.007662i$~\\[0.5ex]
&0.002&$2.743949-0.008586i$~\\[0.5ex]
&0.003&$2.867207-0.009126i$~\\[0.5ex]
&0.004&$2.949297-0.009507i$~\\
[0.5ex]\hline
0.5 &0&$-0.915499+0.001992i$~\\[0.5ex]
&0.001&$1.560090-0.003788i$~\\[0.5ex]
&0.002&$1.943326-0.004890i$~\\[0.5ex]
&0.003&$2.149033-0.005538i$~\\[0.5ex]
&0.004&$2.286064-0.005998i$~\\
[0.5ex]\hline
0.7&0.001&$0.601607-0.001210i$~\\[0.5ex]
&0.002&$1.138613-0.002375i$~\\[0.5ex]
&0.003&$1.426927-0.003055i$~\\[0.5ex]
&0.004&$1.619019-0.003538i$~\\
[0.5ex]\hline
0.9&0.001&$-0.360504+0.000625i$~\\[0.5ex]
&0.002&$0.330434-0.000590i$~\\[0.5ex]
&0.003&$0.701466-0.001283i$~\\[0.5ex]
&0.004&$0.948705-0.001770i$~\\
[0.5ex]\hline
1.1&0.002&$-0.480700+0.000759i$~\\[0.5ex]
&0.003&$-0.026874+0.000043i$~\\[0.5ex]
&0.004&$0.275571-0.000450i$~\\
[0.5ex]\hline
1.3&0.003&$-0.757706+0.001010i$~\\[0.5ex]
&0.004&$-0.400013+0.000586i$~\\
[0.5ex]\hline\hline
\end{tabular}
\end{table}
\end{center}

From now on we concentrate on the Born-Infeld AdS black hole
spacetime.When the Born-Infed parameter tends to zero, the spacetime
boils down to the Reissner-Nordstrom AdS black hole. For the
Reissner-Nordstrom AdS black hole, we know that there is an upper
bound of the electric charge contained inside the black hole,
$Q_{ex}=\sqrt{\frac{3+6 r_h^2}{2}}\frac{ r_h^2 }{\kappa}$. For the
Born-Infeld AdS black hole, this extreme charge $Q_{ex}$ can be
calculated numerically.

When we choose the coupling parameter $q$ between
the scalar field and the electric field to be
zero and the Born-Infeld parameter $b=0$, we can
obtain the QNM for the single scalar perturbation
in the Reissner-Nordstrom AdS background as
listed in table II, which is consistent with the
computation in \cite{wang}. When the
Reissner-Nordstrom AdS black hole becomes more
charged, the real part of the quasinormal
frequency decreases while the imaginary part
becomes more negative.  Keeping $q=0$ and turning
on the Born-Infeld parameter $b$, we can have the
QNM of the single scalar field in the Born-Infeld
AdS black hole as shown in table III. The
nonlinear property of the electric field will not
change the dependence of the quasinormal
frequency on the black hole charge.

\begin{center}\label{table51}
\begin{table}[ht]
\caption{\label{table51} $\kappa=1$, $q=0$, $b=0$. The dependence of
the QNM on Q and $r_h$ around the Reissner-Nordstrom AdS black hole.}
\begin{tabular}
{p{10mm}|p{15mm}|p{12mm}|p{45mm}}\hline\hline
&&&\\ [-2ex]
$~~~r_h$&$~~~~Q$ &$~~Q/Q_{ex}$ &$~~~~~~~~~~~~~\omega$\\
\hline
~~0.1&$~~0.00123$&~~~0.1& $~~~~~~3.934866-0.018082i$~\\
&$~~0.00247$ &~~~0.2&$~~~~~~3.932722-0.018429 i$~\\
&$~~0.00495$ &~~~0.4&$~~~~~~3.923931-0.020024i$~\\
&$~~0.00742$ &~~~0.6 &$~~~~~~3.908544-0.023703i$~\\
&~~0.00990&~~~0.8&$~~~~~~3.886169-0.032180 i $~\\
&~~0.01113&~~~0.9&$~~~~~~3.873996-0.039081i$~\\
&~~0.01175&~~~0.9&$~~~~~~3.867630-0.043017i$~\\ \hline
~~0.2&~~0.00509&~~~0.1&~~~~~~$3.788329-0.168287i$~\\
&~~0.01018&~~~0.2&~~~~~~$3.781288-0.173062i$~\\
&~~0.02036&~~~0.4&~~~~~~$3.753501-0.194535i$~\\
&~~0.03055&~~~0.6&~~~~~~$3.711659-0.239103i$~\\
&~~0.04073&~~~0.8&~~~~~~$3.673163-0.310469i$~\\
&~~0.04582&~~~0.9&~~~~~~$3.658485-0.350673i$~\\
&~~0.04837&~~~0.95&~~~~~~$3.651815-0.371542i$~\\
[0.5ex]\hline\hline
\end{tabular}
\end{table}
\end{center}
\begin{center}\label{table52}
\begin{table}[ht]
\caption{\label{table52} The dependence of the QNM of single scalar field
on the Born-Infeld AdS black hole background. Here we have chosen $b=1$.}
\begin{tabular}{p{10mm}|p{15mm}|p{12mm}|p{45mm}}\hline\hline
&&&\\ [-2ex]
$~~~r_h$&$~~~~Q$ &$~~Q/Q_{ex}$ &$~~~~~~~~~~~~~\omega$\\
\hline
~~0.01&$~~0.00150$&~~~0.1& $~~~~~~3.999782-0.0000123i$~\\
&$~~0.00300$ &~~~0.2&$~~~~~~3.995464-0.0000125i$~\\
&$~~0.00600$ &~~~0.4&$~~~~~~3.989700-0.0000129i$~\\
&$~~0.00900$ &~~~0.6 &$~~~~~~3.983007-0.0000135i$~\\
&~~0.01200&~~~0.8&$~~~~~~3.975621-0.0000146i $~\\
&~~0.01350&~~~0.9&$~~~~~~3.971702-0.0000158i$~\\
&~~0.01425&~~~0.95&$~~~~~~3.969690-0.0000175i$~\\ \hline
~~0.1&~~0.01535&~~~0.1&~~~~~~$3.909637-0.020425i$~\\
&~~0.03070&~~~0.2&~~~~~~$3.868855-0.024228i$~\\
&~~0.06140&~~~0.4&~~~~~~$3.762048-0.037783i$~\\
&~~0.09210&~~~0.6&~~~~~~$3.621570-0.071016i$~\\
&~~0.12280&~~~0.8&~~~~~~$3.460195-0.155326i$~\\
&~~0.13815&~~~0.9&~~~~~~$3.388106-0.217904i$~\\
&~~0.14582&~~~0.95&~~~~~~$3.355319-0.252168i$~\\
[0.5ex]\hline\hline
\end{tabular}
\end{table}
\end{center}

Now we turn on the coupling between the scalar field and the electromagnetic field. In the Reissner-Nordstrom AdS background, when the Born-Infeld parameter is zero, the scalar perturbation coupled with the electromagnetic field was studied in \cite{hexi}. Here we present the property of the QNM of the scalar field with the change of its coupling strength with the electromagnetic field and the electric charge of the Reissner-Nordstrom black hole. Our numerical results are shown in table IV. We find that for the nonzero $q$, the absolute value of the imaginary part of the quasinormal frequency does not keep on increasing with the increase of the black hole charge as we observed when $q=0$. It first decreases and then increases when the black hole becomes more charged. When the strength of the coupling $q$ is big enough, the imaginary frequency will keep on increasing with the increase of the black hole electric charge until it jumps to be positive, showing that the background spacetime becomes unstable. The real part keeps the same property as the case when $q=0$, it continues to decrease with the increase of the black hole charge, except for the big enough $q$ it becomes unphysical when the spacetime is destroyed. This property can be seen clearly in Fig.1.

\begin{center}\label{table60}
\begin{table}[ht]
\caption{\label{table60} The frequencies of the charged scalar
field when we choose the coupling with the Maxwell field $q=0.5,2$ and $4$ respectively around the RN-AdS black hole with
$r_h=0.1$. }
\begin{tabular}{p{10mm}|p{15mm}|p{12mm}|p{45mm}}\hline\hline
&&&\\ [-2ex] $q$&$~~~~Q$ &$~~Q/Q_{ex}$ &$~~~~~~~~~~~~~~~~~~~\omega$\\
\hline &&&\\ [-2ex]
0.5&~~~~~0 &~~~~~0& $~~~~~~~3.935576-0.017971i$~\\
&$~~0.00124$&~~~0.1& $~~~~~~~3.875069-0.017627i$~\\
&$~~0.00247$ &~~~0.2&$~~~~~~~3.813143-0.017503i$~\\
&$~~0.00371$ &~~~0.3&$~~~~~~~3.749777-0.017612i$~\\
&$~~0.00495$ &~~~0.4&$~~~~~~~3.684917-0.017997i$~\\
&$~~0.00618$ &~~~0.5 &$~~~~~~~3.618480-0.018750i$~\\
&$~~0.00742$ &~~~0.6 &$~~~~~~~3.550352-0.020045i$~\\
&$~~0.00866$ &~~~0.7 &$~~~~~~~3.480442-0.022210i$~\\
&~~0.00990&~~~0.8&$~~~~~~~3.408885-0.025790i $~\\
&~~0.01113&~~~0.9&$~~~~~~~3.336418-0.031127i$~\\
&~~0.01175&~~~0.95&$~~~~~~~3.300021-0.034267i$~\\
\hline &&&\\ [-2ex]
2&~~~~~0 &~~~0& $~~~~~~~3.935576-0.017971i$~\\
&$~~0.00124$&~~~0.1& $~~~~~~~3.695650-0.016306i$~\\
&$~~0.00247$ &~~~0.2&$~~~~~~~3.454297-0.014902i$~\\
&$~~0.00371$ &~~~0.3&$~~~~~~~3.211561-0.013708i$~\\
&$~~0.00495$ &~~~0.4&$~~~~~~~2.967456-0.012691i$~\\
&$~~0.00618$ &~~~0.5 &$~~~~~~~2.721970-0.011838i$~\\
&$~~0.00742$ &~~~0.6 &$~~~~~~~2.475050-0.011157i$~\\
&$~~0.00866$ &~~~0.7 &$~~~~~~~2.226596-0.010705i$~\\
&~~0.00990&~~~0.8&$~~~~~~~1.976435-0.010659i $~\\
&~~0.01113&~~~0.9&$~~~~~~~1.724470-0.011570i$~\\
&~~0.01175&~~~0.95&$~~~~~~~1.598092-0.012641i$~\\
\hline &&&\\[-2ex]
4&~~~~~0 &~~~0& $~~~~~~~3.935576-0.017971i$~\\
&$~~0.00124$&~~~0.1& $~~~~~~~3.456359-0.014642i$~\\
&$~~0.00247$ &~~~0.2&$~~~~~~~2.975567-0.011821i$~\\
&$~~0.00371$ &~~~0.3&$~~~~~~~2.493319-0.009374i$~\\
&$~~0.00495$ &~~~0.4&$~~~~~~~2.009714-0.007207i$~\\
&$~~0.00618$ &~~~0.5 &$~~~~~~~1.524838-0.005249i$~\\
&$~~0.00742$ &~~~0.6 &$~~~~~~~1.038765-0.003447i$~\\
&$~~0.00866$ &~~~0.7 &$~~~~~~~0.551568-0.001766i$~\\
&~~0.00990&~~~0.8&$~~~~~~~0.063334-0.000194i $~\\
&~~0.01113&~~~0.9&$~~~-0.425792+0.001168i$~\\
&~~0.01175&~~~0.95&$~~~-0.670674+0.001581i$~\\
[0.5ex]\hline\hline
\end{tabular}
\end{table}
\end{center}

When we consider the electromagnetic field to be nonlinear with nonzero $b$, the dependence of the quasinormal frequencies of the scalar field on the coupling strength with the electromagnetic field and the electric charge $Q$ does not change. The behaviors are shown in Fig.2 for the real part and the imaginary part of the quasinormal frequencies. Comparing with the zero Born-Infeld parameter case, nonzero $b$ makes the turning point of the imaginary part of the quasinormal frequency appear for smaller $Q/Q_{ex}$ when $q$ is fixed and the positive imaginary part of the frequency can appear  for bigger coupling strength $q$. Fixing $q=4$, the influence of the Born-Infeld parameter is shown in Fig.3. Similar to what we have observed in the Schwarzschild AdS background, when the Born-Infeld parameter is smaller, the coupling between the scalar field and the electromagnetic field can make the background become unstable more easily.

\begin{figure}[ht]\label{case4re0}
\includegraphics[width=190pt]{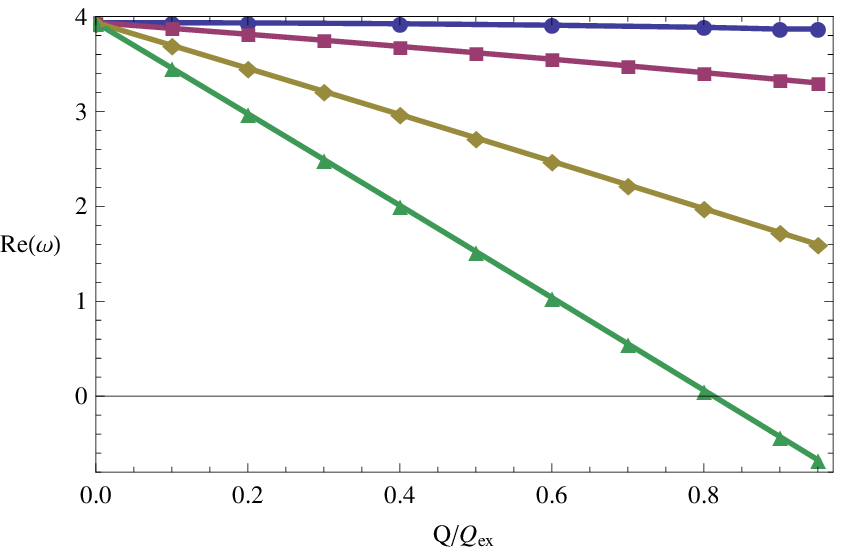}~~~~~~~~
\includegraphics[width=200pt]{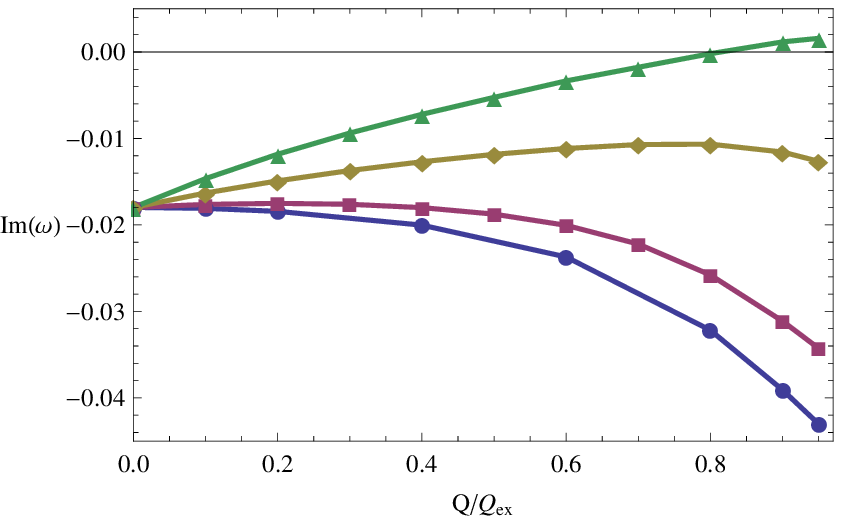}
\caption{\label{case4re0}  The quasinormal frequencies change with the black hole charge in the Reissner-Nordstrom AdS black hole background with the black hole size $r_h=0.1$. For the real part of the frequency, lines from the top to the
bottom correspond to $q=0, 0.5, 2, 4$, respectively.
For the imaginary part of the frequency, lines from the bottom to the top correspond to $q=0, 0.5, 2, 4$, respectively.}
\end{figure}

\begin{figure}[ht]\label{case4re2}
\includegraphics[width=190pt]{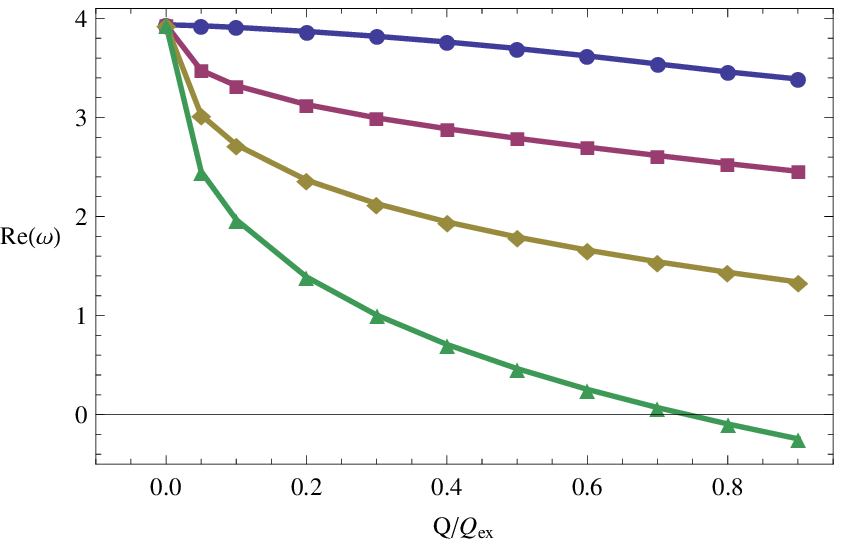}~~~~~~~~
\includegraphics[width=200pt]{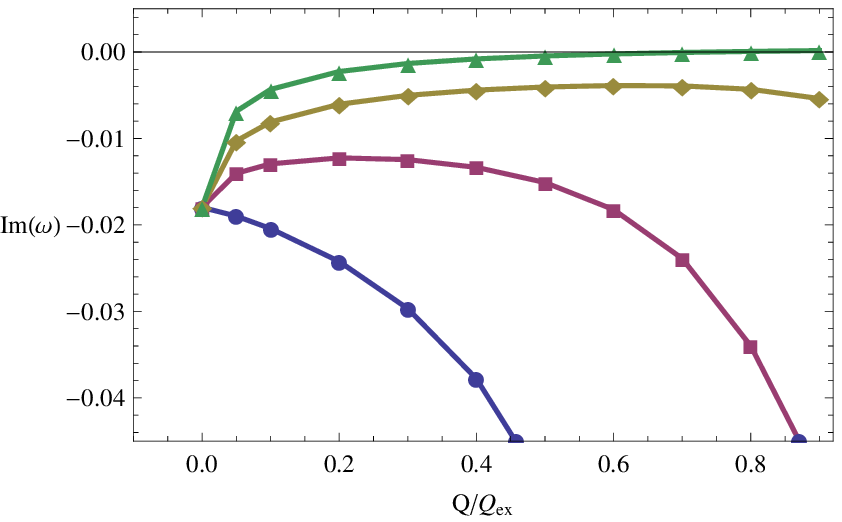}
\caption{\label{case4re2} The quasinormal frequencies change with the black hole charge in the Born-Infeld AdS black hole background with the Born-Infeld parameter $b=1$. The black hole size is fixed as $r_h=0.1$. For the real part of the frequency, lines from the top to the
bottom correspond to $q=0, 2, 4, 6.5$, respectively.
For the imaginary part of the frequency, lines from the bottom to the top correspond to $q=0, 2, 4, 6.5$, respectively.}
\end{figure}

\begin{figure}[h]\label{casenewre1}
\includegraphics[width=190pt]{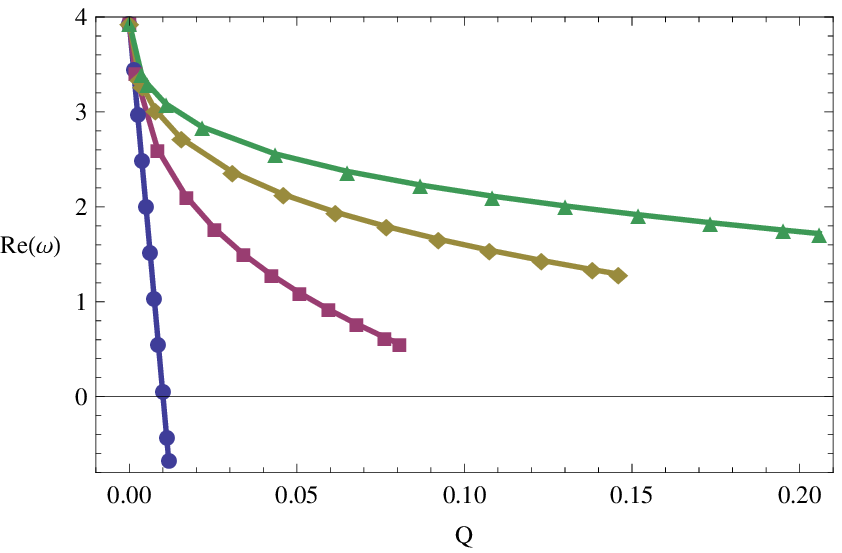}~~~~~~~~
\includegraphics[width=200pt]{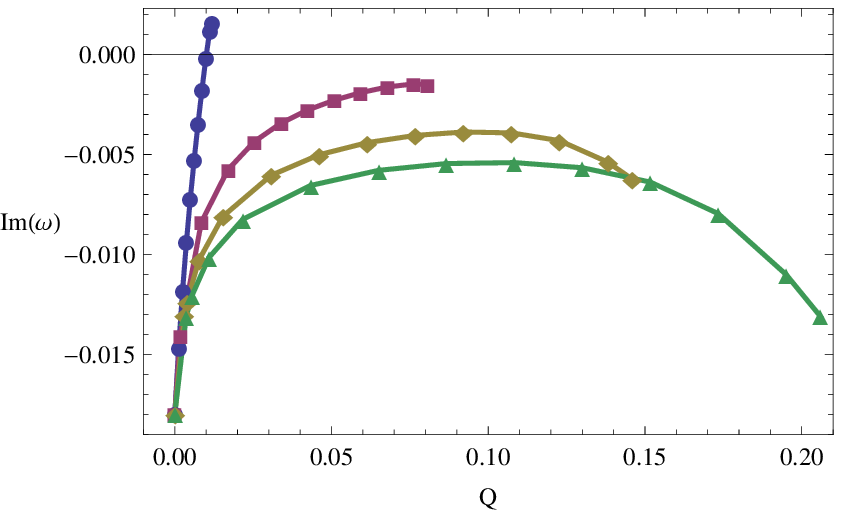}
\caption{\label{casenewre1}
The quasinormal frequencies change with the black hole charge in the AdS black hole background with the $q=4$ fixed and varying Born-Infeld parameter. The black hole size is fixed as $r_h=0.1$. For the real part of the frequency, lines from the top to the
bottom correspond to $b=2, 1, 0.3, 0$, respectively.
For the imaginary part of the frequency, lines from the bottom to the top correspond to $b=2, 1, 0.3, 0$, respectively.
}
\end{figure}

From the above discussion, we learnt that the QNM
of charged scalar field perturbation in the AdS
background is influenced by the electric charge
$Q$, the coupling $q$ between the scalar field
and the electromagnetic field and the Born-Infeld
parameter $b$. To see more clearly the combined
influence, we list our numerical results of the
quasinormal frequencies in table V and table VI
and compare with earlier result presented in
table I. For the chosen value of $q$, we see that
the real part of the frequency keeps increasing
when the Born-Infeld parameter becomes bigger.
The imaginary part of the frequency does not
change monotonously. As shown in table V, the
absolute value of the imaginary frequency
decreases first and then increases when the
Born-Infeld parameter becomes bigger for small
coupling case. When the coupling $q$ becomes
stronger, the turning point of the imaginary
frequency appears for smaller $b$. When $q$ is
big enough, the imaginary frequency becomes
monotonous and appears more negative when the
Born-Infeld parameter is bigger as shown in table
VI. This is consistent with that in table I. From
the third equation in Eq.(\ref{symmetry}), we can
see when $q$ is bigger, the backreaction becomes
weaker. Thus as $q$ increases to be big enough
the behavior should be similar to that in the
probe limit. The property described above is
shown in fig.4.

\begin{center}\label{table6}
\begin{table}[ht]
\caption{\label{table6}  The
dependence of the QNM on $q$ and $b$, where we have chosen $\kappa=1$, $r_h=0.1$, and $Q=0.01$. }
\begin{tabular}{p{10mm}|p{15mm}|p{35mm}|p{10mm}|p{15mm}|p{35mm}}
\hline \hline &&&&&\\ [-2ex]
q& b&$\omega$&q&b&$\omega$~\\[0.5ex]\hline
&&&&&\\ [-2ex]
0.5 &0&$3.402781-0.026176i$~&1 &0&$2.920261-0.020304i$~\\[0.5ex]
&$10^{-4}$&$3.405373-0.026070i$~&&$10^{-4}$&$2.925335-0.020222i$~\\[0.5ex]
&$10^{-2}$&$3.519682-0.021947i$~&&$10^{-2}$&$3.146037-0.017604i$~\\[0.5ex]
&$0.1$&$3.673134-0.018869i$~&&$0.1$&$3.437811-0.016273i$~\\[0.5ex]
&$0.5$&$3.764919-0.017945i$~&&$0.2$&$3.518353-0.016169i$~\\[0.5ex]
&$1$&$3.796554-0.017752i$~&&$0.3$&$3.561588-0.016153i$~\\[0.5ex]
&$10$&$3.870019-0.017575i$~&&$0.5$&$3.611629-0.016171i$~\\[0.5ex]
&$15$&$3.878750-0.017586i$&&$1$&$3.671523-0.016248i$~\\[0.5ex]
&$20$&$3.884328-0.017597i$~&&$5$&$3.777466-0.016567i$~\\[0.5ex]
&$25$&$3.888326-0.017608i$&&$10$&$3.810748-0.016733i$~\\[0.5ex]
&&&&$15$&$3.827325-0.016831i$~\\ [0.5ex]\hline\hline &&&&&\\ [-2ex]
1.5 &0&$2.437763-0.015128i$~&2&0&$1.955207-0.010687i$~\\[0.5ex]
&$10^{-4}$&$2.445310-0.015078i$&&$10^{-4}$&$1.965213-0.010673i$~\\[0.5ex]
&$10^{-2}$&$2.772025-0.013916i$&&$10^{-3}$&$2.042778-0.010604i$~\\[0.5ex]
&$0.1$&$3.202059-0.013993i$&&$10^{-2}$&$2.397601-0.010808i$~\\[0.5ex]
&$0.2$&$3.320658-0.014206i$&&$0.1$&$2.965891-0.011989i$~\\[0.5ex]
&$0.5$&$3.457998-0.014562i$&&$0.3$&$3.206688-0.012739i$~\\[0.5ex]
&$1$&$3.546196-0.014866i$~&&$0.5$&$3.304039-0.013100i$~\\[0.5ex]
&$5$&$3.702263-0.015613i$~&&$1$&$3.420584-0.013594i$~\\[0.5ex]
&$10$&$3.751320-0.015929i$&&$10$&$3.691739-0.015163i$~\\[0.5ex]
&$15$&$3.775763-0.016108i$~&&$20$&$3.744733-0.015586i$~\\
[0.5ex]\hline\hline
\end{tabular}
\end{table}
\end{center}

\begin{center}\label{table7}
\begin{table}[ht]
\caption{\label{table7}  The
dependence of the QNM on $q$ and $b$, where we have chosen $\kappa=1$, $r_h=0.1$ and $Q=0.01$. }
\begin{tabular}{p{10mm}|p{35mm}|p{35mm}|p{35mm}}
\hline \hline &&&\\ [-2ex]
b&$~~~~~~\omega$~~(q=3)&$~~~~~~\omega$ ~~(q=4)&$~~~~~~\omega$ ~~(q=5)~\\[0.5ex]\hline
&&&\\ [-2ex]
0&$0.989457-0.004056i$&$0.022024-0.000067i$~&$-0.947650+0.002231i$~\\[0.5ex]
$10^{-4}$&$1.004346-0.004101i$&$0.041779-0.000127i$~&$-0.923010+0.002183i$~\\[0.5ex]
$10^{-2}$&$1.647412-0.006033i$&$0.895384-0.002710i$~&$0.141586-0.000363i$~\\[0.5ex]
$0.1$&$2.492360-0.008681i$&$2.017336-0.006116i$~&$1.540944-0.004112i$~\\[0.5ex]
$0.5$&$2.975184-0.010568i$&$2.685157-0.008475$~&$2.374046-0.006739i$~\\[0.5ex]
$1$&$3.168540-0.011349i$&$2.915465-0.009446i$~&$2.661429-0.007831i$~\\[0.5ex]
5&$3.475528-0.013057i$&$3.323476-0.011580i$~&$3.170748-0.010263i$~\\[0.5ex]
$10$&$3.572124-0.013735i$&$3.451928-0.012436i$~&$3.331173-0.011255i$~\\[0.5ex]
$20$&$3.651069-0.014369i$&$3.556941-0.013243i$~&$3.462364-0.012202i$~\\
[0.5ex]
$40$&$3.714917-0.014953i$&$3.641898-0.013993i$&$3.568529-0.013091i$~\\
[0.5ex]
$80$&$3.766023-0.015482i$&$3.709919-0.014677i$&$3.653550-0.013911i$~\\
[0.5ex]
$100$&$3.780121-0.015640i$&$3.728684-0.014882i$&$3.677008-0.014158i$\\
[0.5ex]\hline\hline
\end{tabular}
\end{table}
\end{center}

\begin{figure}[ht]\label{case4re1}
\includegraphics[width=190pt]{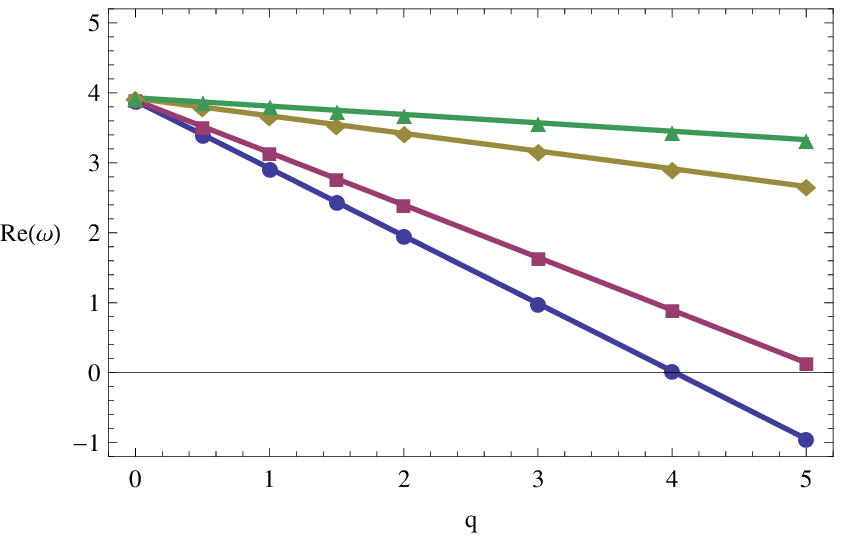}~~~~~~~~
\includegraphics[width=200pt]{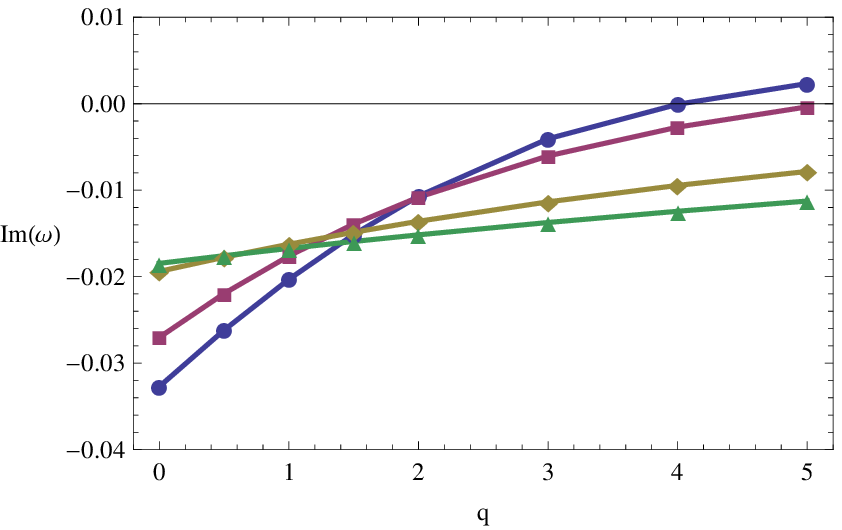}
\caption{\label{case4re1} The quasinormal
frequencies change with the coupling parameter
$q$ in the AdS black hole background for
different Born-Infeld parameter $b$. The black
hole size is fixed as $r_h=0.1$ and the electric
charge is chosen as $Q=0.01$. For the real part
of the frequency, lines from the top to the
bottom correspond to $b=10, 1, 0.01, 0$,
respectively. For the imaginary part of the
frequency, lines from the bottom to the top
correspond to $b=10, 1, 0.01, 0$, respectively. }
\end{figure}

\section{Summary}

In this work we have studied the QNM of the perturbation of the scalar field interacting with the electromagnetic field in the backgrounds of neutral AdS black hole, the Reissner-Nordstrom AdS black hole and the Born-Infeld AdS black hole. We have disclosed rich physics brought by the electric charge $Q$, the coupling between the scalar and electromagnetic field $q$ and the Born-Infeld parameter $b$ on the quasinormal frequencies of the scalar perturbation. Different from the single classical field perturbation, which always has the decay mode in the black hole background, we observed the possible growing mode when the perturbation of the scalar field strongly couples to the electromagnetic field. This indicates that the interaction among classical fields can destroy the background spacetime. In the language of the AdS/CFT, this is due to the condensate of the scalar field on the black hole background and there is a second order transition between a non-superconducting state at high temperatures and a superconducting state at low temperatures. Our results disclose the signature of how this phase transition happens from the phenomenon of the perturbation. When the electromagnetic field becomes nonlinear described by the Born-Infeld electrodynamics, the nonlinearity described by the Born-Infeld parameter can hinder the instability to happen in the scalar perturbation around the background spacetime. The QNM behavior of the scalar field disclosed here can help us further understand the physics in the holographic superconductor.

\emph{Acknowledgment: }
 We thank Qiyuan Pan and
Ma Lei for useful discussions. This work was
supported in part by NNSF of China.

\end{document}